\def\be{\begin{equation}}
\def\ee{\end{equation}}
\def\ba{\begin{array}}
\def\ea{\end{array}}
\def\Tr{\mathrm{Tr}}

\documentclass[groupedaddress,aps,pra,superscriptaddress,showpacs,twocolumn,prl]{revtex4}%
\usepackage{epsfig,dsfont,amssymb,amsmath,amsthm,amsfonts,amsbsy,mathrsfs}
\usepackage{graphicx, color}
\usepackage{amsmath}
\usepackage{amssymb}
\usepackage{mathdots}
\usepackage{float}
\usepackage{cases}
\usepackage{graphpap}
\usepackage{wrapfig}
\usepackage{tikz}
\usepackage{indentfirst}
\usepackage{ctex}
\usepackage{picinpar,graphicx}
\begin{document}
\title{Strong entropic uncertainty relations for multiple measurements}
\author{Yunlong Xiao}
\affiliation{School of Mathematics, South China University of Technology, Guangzhou, Guangdong 510640, China}
\affiliation{Max Planck Institute for Mathematics in the Sciences, 04103 Leipzig, Germany}
\author{Naihuan Jing}
\thanks{Corresponding author: jing@ncsu.edu}
\affiliation{Department of Mathematics, North Carolina State University, Raleigh, NC 27695, USA}
\affiliation{School of Mathematics, South China University of Technology, Guangzhou, Guangdong 510640, China}
\author{Shao-Ming Fei}
\affiliation{School of Mathematical Sciences, Capital Normal University, Beijing 100048, China}
\affiliation{Max Planck Institute for Mathematics in the Sciences, 04103 Leipzig, Germany}
\author{Tao Li}
\affiliation{School of Science, Beijing Technology and Business University,  Beijing 102488,  China}
\author{Xianqing Li-Jost}
\affiliation{Max Planck Institute for Mathematics in the Sciences, 04103 Leipzig, Germany}
\author{Teng Ma}
\affiliation{School of Mathematical Sciences,  Capital Normal University,  Beijing 100048,  China}
\author{Zhixi Wang}
\affiliation{School of Mathematical Sciences,  Capital Normal University,  Beijing 100048,  China}

\begin{abstract}
 In this paper, we study entropic uncertainty relations on a finite-dimensional Hilbert space and provide several tighter bounds for multi-measurements,
 with some of them also valid for R\'{e}nyi and Tsallis entropies besides the Shannon entropy. We employ majorization theory and actions of the symmetric group to obtain an {\it admixture bound} for
 entropic uncertainty relations for multi-measurements.
 Comparisons among all bounds for multi-measurements are shown in figures in our favor.
\end{abstract}


\maketitle

\section{\uppercase\expandafter{\romannumeral1}. Introduction}
The most revolutionary departure of quantum mechanics from classical mechanics
is that it is impossible to simultaneously measure two complementary variables of a particle in precision. Kennard's form of the
Heisenberg uncertainty principle \cite{Heisenberg} displays vividly such an inequality for the standard deviation of
position and momentum of a particle:
$\sigma_{Q}\sigma_{P}\geqslant\frac{1}{2}$,
where the Planck constant is taken as $\hbar=1$.
The corresponding entropic uncertainty of
Bia{\l}ynicki-Birula and Mycielski \cite{Bialynicki} says that
$h(Q)+h(P)\geqslant\log(e\pi)$,
where $Q$ and $P$ stand for position and momentum respectively while $h$ is the differential entropy: $h(Q)=-\int_{-\infty}^{\infty}f(x)\log f(x)dx$
with $f(x)$ being the probability density corresponding to $Q$.

In the seminal paper \cite{Deutsch}, Deutsch studied the entropic uncertainty relations on finite $d$-dimensional Hilbert spaces
in terms of the Shannon entropy for any two measurements $M_{1}$ and $M_{2}$ (base $2$ $\log$ is used unless stated otherwise):
\begin{align}\label{e:D}
H(M_{1})+H(M_{2})\geqslant-2\log\frac{1+\sqrt{c_{1}}}{2},
\end{align}
where $c_{1}$ is the largest element in the overlap matrix $c(M_1, M_2)$ of the two measurements.
Later Maassen and Uffink \cite{Maassen, Berta} derived the influential generalized quantum mechanical uncertainty
relation which amounts to a tighter lower bound than Eq. (\ref{e:D}). Recently
Coles and Piani \cite{Coles} proved that,
for any two measurements $M_j=\{|u_{i_j}^j\rangle\}$ on a quantum state $\rho$ over a finite dimensional Hilbert space
\begin{align}\label{e:cp1}
H(M_{1})+H(M_{2})\geqslant-\log c_{1}+\frac{1-\sqrt{c_{1}}}{2}\log\frac{c_{1}}{c_{2}},
\end{align}
where $c_{2}$ is the second largest value among all overlaps $c(u^{1}_{i_{1}}, u^{2}_{i_{2}})=|\langle u^{1}_{i_{1}}|u^{2}_{i_{2}}\rangle|^{2}$.
Then Maassen-Uffink's bound is simply obtained by dropping the second term in RHS of Eq. (\ref{e:cp1}).

More recently, S. Liu {\it et al.} \cite{Fan} generalized Coles and Piani's method to give a lower bound for $N$ measurements $M_i$:
\begin{align}\label{e:Liu}
\sum\limits_{m=1}^{N}H(M_{m})\geqslant-\log b+(N-1)S(\rho),
\end{align}
where
\begin{align}
b=\max\limits_{i_{N}} \left\{\sum\limits_{i_{2}\sim i_{N-1}} \max\limits_{i_{1}} [c(u^{1}_{i_{1}}, u^{2}_{i_{2}})] \prod\limits_{m=2}^{N-1} c(u^{m}_{i_{m}}, u^{m+1}_{i_{m+1}}) \right\}.
\end{align}
and $S(\rho)$ is the {\it von Neumann entropy} of the quantum state $\rho$. Thus the
state-independent uncertainty relation for multi-measurement is the corresponding inequality by ignoring $S(\rho)$.
In fact, the state-independent inequality generalizes Maassen-Uffink's bound, which suggests that there are rooms for improvement
in regards to Coles-Piani's bound. Such an improvement will be useful for further applications in quantum information processing, especially in
quantum cryptography when several measurements are present. For the importance of entropic uncertainty relations and other applications,
the reader is referred to \cite{Tomamichel2, Tomamichel}.

The aim of this article is to find several tighter bounds for multi-measurements in comparison with the bound of Eq.(\ref{e:Liu}) by using majorization theory and symmetry.
Of course it is a combinatorial or mathematical exercise to obtain
bounds for multi-measurements based on the usual entropic sum of two measurements. However, what we will show
is that deeper analysis is needed for nontrivial and tighter bounds for multi-measurements, and
applications of majorization theory and symmetry inside the physical construction help to obtain true generalization for multi-measurements.

Indeed, from the construction of the universal uncertainty relation \cite{Friedland, Puchala}, the joint probability distribution in vector
$P^{1}\otimes P^{2}$, with respect to the measurement $M_{1}$ and $M_{2}$, should be controlled by a bound $\omega$
that quantifies its uncertainty in terms of majorization and is also independent of the state $\rho$. Thus, $H(P^{1})+H(P^{2})\geqslant H(\omega)$
for any nonnegative Schur concave function $H$ such as the Shannon entropy.
Therefore, the generalized universal uncertainty relation for $N$ measurements
$$\bigotimes\limits_{m=1}^{N} P^{m}\prec\omega$$
can imply that $\sum\limits_{m=1}^{N}H(P^{m})\geqslant H(\omega)$ for multi-measurements.
In section \uppercase\expandafter{\romannumeral2}, we first give a precise formula of majorization bound for $N$ probability distributions,
and discuss two simple forms of the majorization bounds for multi-measurements in connection with Eq. (\ref{e:Liu}).
Comparison of our bounds with previously ones in figure 1 shows that our bounds are tighter.

Further study shows that the simple sum of the uncertainties does not completely reveal
the physical meaning of the entropic bounds. The reason is that when one computes the sum of the entropies such as Eq. (\ref{e:Liu}),
the mathematical
summation does not really provide physically cor\-rect answer, as the measurement outcomes clearly do not know which order
we perform the measurements, and the bound for $N$-measurement should be independent from the order of measuring.
Therefore one should consider the average of all possible orders of measurements. But this
average is cumbersome and does not provide good enough result.

In order to solve this and get operational formulas for the entropic uncertainty relation of multi-measurements, we
study the effects of symmetry on majorization bounds in section \uppercase\expandafter{\romannumeral3}
and find that there is a large invariant subgroup of the full symmetry group under the action on
certain products of probability distribution vectors and
logarithms of remaining distributions. After factoring out this invariant factor we
obtain a simple average to give our main result in Section \uppercase\expandafter{\romannumeral3}:
\begin{align}
\sum\limits_{m=1}^{N}H(M_{m})+(1-N)S(\rho)\geqslant-\frac{1}{N}\omega\mathfrak{B}.  
\end{align}
where $\omega$ is the universal majorization bound of $N$-measurements and $\mathfrak B$ is certain vector of logarithmic distributions
(cf. Theorem 3).
We call this bound an {\it admixture bound}, since it is obtained by mixing the universal bound from tensor products and factoring out the action of
the invariant subgroup of the symmetric
group. We then show that this admixture bound is tighter than all previously known bounds in the last part of the section.
The exact comparison is charted in figure 2.


\section{\uppercase\expandafter{\romannumeral2}. Universal bounds of Majorization}
Majorization characterizes a balanced partial relationship between two vectors that are comparable and was
 studied long ago in algebra and analysis. It has been used to study entropic uncertainty relations
\cite{Partovi, Hossein} and played an important role in formulation of state-independent entropic uncertainty relations
\cite{Friedland, Puchala, Rudnicki}.
A vector $x$ is majorized by another vector $y$ in $\mathbb R^d$ : $x\prec y$ if
$\sum\limits_{i=1}^{k}x_{i}^{\downarrow}\leqslant\sum\limits_{i=1}^{k}y_{i}^{\downarrow} (k=1, 2, \cdots, d-1)$
and $\sum\limits_{i=1}^{d}x_{i}^{\downarrow}=\sum\limits_{i=1}^{d}y_{i}^{\downarrow}$,
where the down-arrow denotes that the components are
ordered in decreasing order $x_{1}^{\downarrow}\geqslant \cdots \geqslant x_{d}^{\downarrow}$.
A nonnegative Schur concave function $\Phi$ on $\mathbb R^d$ preserves the partial order in the sense that
$x\prec y$ implies that $\Phi(x)\geqslant \Phi(y)$. We adopt the convention to write a probability distribution
vector in a short form by omitting the string of zeroes at the end, for example, $(0.6, 0.4, 0, \cdots, 0)=(0.6, 0.4)$
and the actual dimension of the vector should be clear from the context.

The tensor product $x\otimes y$ of two vectors $x=(x_1, \cdots, x_{d_1})$ and $y=(y_1, \cdots, y_{d_2})$
is defined as $(x_1y_1, \cdots, x_1y_{d_2}, \cdots, x_{d_1}y_1, \cdots, x_{d_1}y_{d_2})$, and multi-tensors
are defined by associativity.
It is well-known that Shannon, R\'{e}nyi
and Tsallis entropies are nonnegative Schur-concave, thus for probability distributions $P^{1}$ and $P^{2}$ with $P^{1}\otimes P^{2}\prec\omega$ implies that
$\Phi(P^{1}\otimes P^{2})\geqslant\Phi(\omega)$ for any of the entropies $\Phi$.

A majorization uncertainty relation for two measurements was well studied in \cite{Friedland, Puchala}.
We now construct the analogous universal upper bound for multi-measurements.
Let $\rho$ be a mixed quantum state on a $d$-dimensional Hilbert space $\mathcal{H}\cong \mathds{C}^{d}$, and
let $M_{m}$ $(m=1, 2, \cdots, N)$ be $N$ measurements. Assume that $M_{m}$ has a set of orthonormal eigenvectors $\{|u^{m}_{i_{m}}\rangle\}$ $(i_m=1, 2, \cdots, d)$, and denote by $P^m=(p^{m}_{i_{m}})$, where $p^{m}_{i_{m}}=\langle u^{m}_{i_{m}}|\rho|u^{m}_{i_{m}}\rangle$ the probability distributions obtained by measuring $\rho$ with respect to bases $\{|u^{m}_{i_{m}}\rangle\}$. We can derive a state-independent bound of $\bigotimes_{m} P^{m}$ under majorization
\begin{align}
\bigotimes\limits_{m=1}^{N} P^{m}\prec\omega,
\end{align}
where the quantity on the left-hand side represents the joint probability distribution induced by measuring $\rho$ with measurements $M_{m}$ $(m=1, 2, \cdots, N)$.

For subsets $\{|u^{1}_{i_{1}}\rangle, \cdots, |u^{1}_{i_{S_{1}}}\rangle\}$, $\{|u^{2}_{j_{1}}\rangle, \cdots, |u^{2}_{j_{S_{2}}}\rangle\}$, $\cdots$, $\{|u^{N}_{l_{1}}\rangle, \cdots, |u^{N}_{l_{S_{N}}}\rangle\}$ of the orthonomal bases of $M^1, M^2, \cdots M^N$ respectively
such that $S_{1}+S_{2}+\cdots+S_{N}=k+N-1$, we define the matrices $U_{ij}(S_i, S_j)$
\begin{align}
&U_{12}(S_{1}, S_{2})=
\left(
\begin{array}{c}
  \langle u^{1}_{i_{1}}| \\
  \langle u^{1}_{i_{2}}| \\
  \vdots  \\
  \langle u^{1}_{i_{S_{1}}}| \\
\end{array}
\right)\cdot\left(|u^{2}_{j_{1}}\rangle, |u^{2}_{j_{2}}\rangle, \cdots, |u^{2}_{j_{S_{2}}}\rangle\right)\notag\\
&=\left(
\begin{array}{cccc}
  \langle u^{1}_{i_{1}}|u^{2}_{j_{1}}\rangle & \langle u^{1}_{i_{1}}|u^{2}_{j_{2}}\rangle & \cdots & \langle u^{1}_{i_{1}}|u^{2}_{j_{S_{2}}}\rangle \\
  \langle u^{1}_{i_{2}}|u^{2}_{j_{1}}\rangle & \langle u^{1}_{i_{2}}|u^{2}_{j_{2}}\rangle & \cdots & \langle u^{1}_{i_{2}}|u^{2}_{j_{S_{2}}}\rangle \\
  \vdots & \vdots & \ddots & \vdots \\
  \langle u^{1}_{i_{S_{1}}}|u^{2}_{j_{1}}\rangle & \langle u^{1}_{i_{S_{1}}}|u^{2}_{j_{2}}\rangle & \cdots & \langle u^{1}_{i_{S_{1}}}|u^{2}_{j_{S_{2}}}\rangle \\
\end{array}
\right).
\end{align}
For simplicity we abbreviate $U_{12}(S_{1}, S_{2})$ by $U_{12}$. Then $U_{13}$, $U_{14}, \cdots U_{N-1, N}$ are constructed similarly.
We define the block matrix
\begin{align}
U(S_{1}, S_{2}, \cdots, S_{N})=
\left(
\begin{array}{cccc}
  I_{S_{1}} & U_{12} & \cdots & U_{1N} \\
  U_{21} & I_{S_{2}} & \cdots & U_{2N} \\
  \vdots & \vdots & \ddots & \vdots \\
  U_{N1} & U_{N2} & \cdots & I_{S_{N}} \\
\end{array}
\right).
\end{align}
Since the eigenvalues of a Hermitian matrix are real, we adopt the convention to label the eigenvalues in decreasing order. Let $\lambda_{1}(\bullet)$ and $\sigma_{1}(\bullet)$ denote the maximal eigenvalue and singular value of a matrix respectively.
Generalizing the idea of \cite{Puchala, Rudnicki}, we introduce the elements $s_{k}$ by
\begin{align}\label{e:sk}
s_{k}=\max\limits_{\sum\limits_{x=1}^{N}S_{x}=k+N-1}\{\lambda_{1}(U(S_{1}, S_{2}, \cdots, S_{N}))\}.
\end{align}
We remark that when $N=2$, Eq. (\ref{e:sk}) will degenerate to the $s_{k}$ defined in \cite{Puchala}. Write
\begin{align}
\Omega_{k}=(\frac{s_{k}}{N})^N,
\end{align}
then we have $\Omega_{1}\leqslant \Omega_{2}\leqslant\cdots \leqslant\Omega_{a}<1$ for some integer $a\leqslant d^N-1$ with $\Omega_{a+1}=1$. With this
preparation we can state our universal upper bound for
multi-measurements:

\noindent\textbf{Theorem 1.} {\it For any $d$-dimensional quantum state $\rho$ and $N$ measurements $M_m$ with their probability distributions $P^{m}$, we have
\begin{align}
\bigotimes\limits_{m=1}^{N} P^{m}\prec\omega,
\end{align}
where
\begin{align}
\omega=(\Omega_{1}, \Omega_{2}-\Omega_{1}, \cdots, 1-\Omega_{a}).
\end{align}
with $a$ being the smallest index such that $\Omega_{a+1}=1$. Here we have used the short form of the $d^N$-dimensional vector $\omega$}.

Theorem 1 is a generalization of the majorization bound for a pair of two measurements \cite{Friedland, Puchala}. Due to its
key role in our discussion, we include a detailed proof.

\noindent\textbf{Proof of Theorem 1}. Consider
sums of $k$ elements from the vector $\bigotimes\limits_{m=1}^{N} P^{m}$, then they are bounded as follows.
\begin{align}
&(p^{1}_{i_{1}}p^{2}_{j_{1}}\cdots p^{N}_{l_{1}})+\cdots+(p^{1}_{i_{k}}p^{2}_{j_{k}}\cdots p^{N}_{l_{k}})\notag\\
\leqslant&\max\limits_{S_{1}+\cdots+S_{N}=k+N-1}(\sum\limits_{x=1}^{S_{1}}\widetilde{p}^{1}_{x})(\sum\limits_{x=1}^{S_{2}}\widetilde{p}^{2}_{x})
\cdots(\sum\limits_{x=1}^{S_{N}}\widetilde{p}^{N}_{x}),
\end{align}
where $\widetilde{p}^{i}_{1}$, $\widetilde{p}^{i}_{2}$, $\cdots$, $\widetilde{p}^{i}_{S_{i}}$ are the greatest $S_{i}$ elements of $p^{i}_{x}$.

Since the arithmetic mean is at least as large as the geometric mean, we derive that
\begin{align}
(\sum\limits_{x=1}^{S_{1}}\widetilde{p}^{1}_{x})(\sum\limits_{x=1}^{S_{2}}\widetilde{p}^{2}_{x})
\cdots(\sum\limits_{x=1}^{S_{N}}\widetilde{p}^{N}_{x})\leqslant
(\frac{\sum\limits_{x=1}^{S_{1}}\widetilde{p}^{1}_{x}+\cdots+\sum\limits_{x=1}^{S_{N}}\widetilde{p}^{N}_{x}}{N})^{N},
\end{align}
On the other hand,
\begin{align}
&\sum\limits_{x=1}^{S_{1}}\widetilde{p}^{1}_{x}+\cdots+\sum\limits_{x=1}^{S_{N}}\widetilde{p}^{N}_{x}\notag\\
\leqslant&\max\limits_{\sum\limits_{x=1}^{N}S_{x}=k+N-1}\{\lambda_{1}(U(S_{1}, S_{2}, \cdots, S_{N}))\}=s_{k},
\end{align}
so we finally get the following estimate:
\begin{align}
\bigotimes\limits_{m=1}^{N} P^{m}\prec(\Omega_{1}, \Omega_{2}-\Omega_{1}, \cdots, 1-\Omega_{a}),
\end{align}
where $\Omega_{k}=(\frac{s_{k}}{N})^N$ and $\Omega_{a+1}$ is the first component equal to $1$, and this gives the desired majorization bound for multi-measurements.

In the case of higher dimensional quantum state $\rho$, $\lambda_{1}(U(S_{1}, S_{2}, \cdots, S_{N}))$ becomes hard to calculate. However, one can
approximate $\lambda_{1}(U(S_{1}, S_{2}, \cdots, S_{N}))$ by the numerical calculation
\begin{align}\label{e:RRR}
\lambda_{1}(U(S_{1}, S_{2}, \cdots, S_{N}))=\max\limits_{|u\rangle} \langle u|U(S_{1}, S_{2}, \cdots, S_{N})|u\rangle,
\end{align}
where the maximum runs over unit vectors $|u\rangle$, then the right-hand side of Eq. (\ref{e:RRR}) is a deformation of the well-known {\it Rayleigh-Ritz ratio}.
As the unit ball formed by the vectors is compact, {\it Weierstra{\ss} Theorem} ensures the existence of $\lambda_{1}$. Here we will give two simple estimates
of the majorization bound for multi-measurements.  To give the first simple estimation, define $CU(1,2)$ as
\begin{align}
CU(S_{1},S_{2})=
\left(
\begin{array}{cccc}
  0 & U_{12} & \cdots & 0 \\
  U_{21} & 0 & \cdots & 0 \\
  \vdots & \vdots & \ddots & \vdots \\
  0 & 0 & \cdots & 0 \\
\end{array}
\right).
\end{align}
Similarly, we can define $CU(S_{i}, S_{j})$ for any pair of $i$, $j$ such that $1\leqslant i, j\leqslant d$. Then
\begin{align}
U(S_{1}, S_{2}, \cdots, S_{N})=&I_{N+k-1}+CU(S_{1}, S_{2})+\cdots\notag\\
+&CU(S_{N-1}, S_{N}).
\end{align}
Using {\it Weyl's Theorem} on eigenvalues of hermitian matrices, we get that
\begin{align}
&\lambda_{1}(U(S_{1}, S_{2}, \cdots, S_{N}))\notag\\
=&1+\lambda_{1}(CU(S_{1}, S_{2})+\cdots+ CU(S_{N-1}, S_{N}))\notag\\
\leqslant&1+\lambda_{1}(CU(S_{1}, S_{2}))+\cdots+\lambda_{1}(CU(S_{N-1}, S_{N}))\notag\\
=&1+\sigma_{1}(U_{12})+\cdots+\sigma_{1}(U_{N-1, N}),
\end{align}
then we define $\widehat{\Omega}_{k}$ by
\begin{align}
\widehat{\Omega}_{k}=(\frac{1+\widehat{s_{k}}}{N})^N,
\end{align}
where
\begin{align}
\widehat{s}_{k}=\max\limits_{\sum\limits_{x=1}^{N}S_{x}=k+N-1}\{\sigma_{1}(U_{12})+\cdots+\sigma_{1}(U_{N-1, N})\}.
\end{align}
Therefore we arrive at the following result.

\noindent\textbf{Theorem 2.} {\it For any $d$-dimensional quantum state $\rho$ and the probability distributions $P^{m}$
associated to $N$ measurements ${M_{m}}$, we have that
\begin{align}
\bigotimes\limits_{m=1}^{N} P^{m}\prec\widehat{\omega},
\end{align}
}

\begin{figure}
\centering
\includegraphics[width=0.4\textwidth]{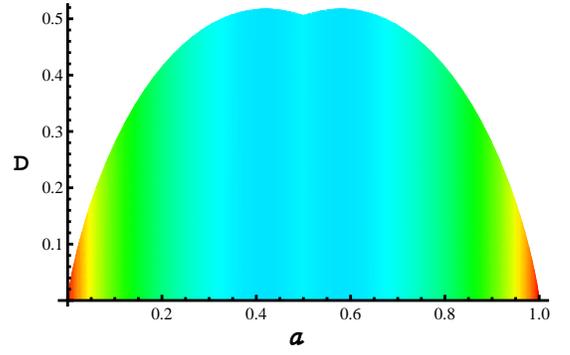}
\caption{Difference (D) of $\log \frac1b$ from $H(\Omega_{1}, 1-\Omega_{1})$ for $\phi=\pi/2$ with respect to $a$. The upper curve
shows the value of $H(\Omega_{1}, 1-\Omega_{1})+\log b$ and it is always nonnegative over $0\leqslant a\leqslant1$.}
\end{figure}

It is obvious from the construction of $\widehat{\omega}$ that the bound is weaker than that of Theorem 1:
$\omega\prec\widehat{\omega}$.

As for the second approximation, note that the universal bound $\omega\prec (\Omega_{1}, 1-\Omega_{1})$, which therefore serves as a simple approximation of $\omega$
for general $N$ probability distributions.
Yet even the bound given by $H(\omega_0)$
with $\omega_0=(\Omega_{1}, 1-\Omega_{1})$ outperforms $-\log b$ appeared in Eq. (\ref{e:Liu}). For example, consider three measurements $M_{i}$ $(i=1, 2, 3)$ in a three-dimensional Hilbert space
 with eigenvectors $u^{1}_{1}=(1, 0, 0)$, $u^{1}_{2}=(0, 1, 0)$, $u^{1}_{3}=(0, 0, 1)$; $u^{2}_{1}=(\frac{1}{\sqrt{2}}, 0, -\frac{1}{\sqrt{2}})$, $u^{2}_{2}=(0, 1, 0)$, $u^{2}_{3}=(\frac{1}{\sqrt{2}}, 0, \frac{1}{\sqrt{2}})$;
$u^{3}_{1}=(\sqrt{a}, e^{i\phi}\sqrt{1-a}, 0)$, $u^{3}_{2}=(\sqrt{1-a}, -e^{i\phi}\sqrt{a}, 0)$ and $u^{3}_{3}=(0, 0, 1)$. With the choice of
$\phi=\pi/2$, we see that the simplest majorization bound $\omega_0=(\Omega_{1}, 1-\Omega_{1})$ under the Shannon entropy is
superior to $-\log b$ over the whole range $0\leqslant a\leqslant1$, where $b=\max_{i_{3}}\{\sum_{i_{2}}
\max_{i_{1}}[c(u^{1}_{i_{1}}, u^{2}_{i_{2}})]c(u^{2}_{i_{2}}, u^{3}_{i_{3}})\}$.
The difference between our second estimation bound $H(\omega_{0})$ and Eq. (\ref{e:Liu}), namely $H(\Omega_{1}, 1-\Omega_{1})+\log b$, is shown in  FIG. 1.

\section{\uppercase\expandafter{\romannumeral3}. Admixture bounds via Symmetry}
As we discussed in the introduction, using Coles and Piani's method, S. Liu {\it et al.} have given an entropic uncertainty bound
for multi-measurements by quantum channels \cite{Fan}:
\begin{align}\label{e:fan}
\sum\limits_{m=1}^{N}H(M_{m})\geqslant-\log b+(N-1)S(\rho),
\end{align}
where
\begin{align}\label{e:b}
b=\max\limits_{i_{N}} \left\{\sum\limits_{i_{2}\sim i_{N-1}} \max\limits_{i_{1}} [c(u^{1}_{i_{1}}, u^{2}_{i_{2}})] \prod\limits_{m=2}^{N-1} c(u^{m}_{i_{m}}, u^{m+1}_{i_{m+1}}) \right\}.
\end{align}

We now use the method of symmetry to significantly strengthen the bound. We note that the above bound depends on the
order of the measurements, so it is natural to denote the bound as $b(M_1, M_2, \cdots, M_N)$ or simply $b(1, 2, \cdots, N)$
to specify the order of the measurements $M_1, \cdots, M_N$.
Using the apparent symmetry of the measurements, we can
define the action of the symmetric group on the bounds. For each permutation $\alpha\in\mathfrak S_N$ we define
\begin{align}
\alpha b(1, \cdots, N)=b(\alpha(1), \alpha(2), \cdots, \alpha(N)).
\end{align}
and observe that $\mathfrak S_N$ leaves the second term $(N-1)S(\rho)$ of Eq. (\ref{e:fan}) invariant.
This immediately implies 
the following entropic uncertainty relation:
\begin{align}
\sum\limits_{m=1}^{N}H(M_{m})+(1-N)S(\rho)\geqslant-\log b_{min},
\end{align}
where
\begin{align}
b_{min}=\min_{\alpha\in \mathfrak S_N}\{b(\alpha(1), \alpha(2), \cdots, \alpha(N))\}.
\end{align}

Apparently $-\log b_{min}\geqslant -\log b$, so this new bound
$-\log b_{min}+(N-1)S(\rho)$ is tighter than the bound appeared in \cite{Fan}.
This shows that the action of the symmetry group can significantly improve the bound.
We remark that a similar consideration has been discussed in \cite{Zhang}. Our treatment has clarified how the symmetric group acts
on the measurements, which plays an important role in our further investigation.

Now we discuss how to blend the $\mathfrak S_N$-symmetry and the method of quantum channels to derive
a tighter bound than we did in the above.

Suppose we are given $N$ measurements $M_1, \cdots, M_N$ with orthonormal bases $\{|u_{i_j}^j\rangle\}$. For a multi-index
$(i_1, \cdots, i_N)$, where $1\leqslant i_j\leqslant d$, we define the
multi-overlap
\begin{align*}
c_{i_1, \cdots, i_N}^{1,\cdots, N}=c(u^{1}_{i_{1}}, u^{2}_{i_{2}})c(u^{2}_{i_{2}}, u^{3}_{i_{3}})\cdots c(u^{N-1}_{i_{N-1}}, u^{N}_{i_{N}}).
\end{align*}
Then we have that (cf. \cite{Fan})
\begin{align}\label{e:basic}
&(1-N)S(\rho)+\sum\limits_{m=1}^{N} H(M_{m})\notag\\
\geqslant&-\Tr(\rho \log \sum\limits_{i_{1}, i_{2}, \cdots, i_{N}} p^{1}_{i_{1}}c_{i_1, \cdots, i_N}^{1,\cdots, N}[u^{N}_{i_{N}}])\notag\\
=&-\sum\limits_{i_{N}}p^{N}_{i_{N}}\log \sum\limits_{i_{1}, i_{2}, \cdots, i_{N-1}} p^{1}_{i_{1}}c_{i_1, \cdots, i_N}^{1,\cdots, N}\notag\\
:=&I(1, 2, \cdots, N),
\end{align}
where $[u]$ stands for $|u\rangle\langle u|$. Note that the above inequality is obtained by a fixing order of $M_1, \cdots, M_N$ which
explains why we can denote the last expression as $I(1, 2, \cdots, N)$.
Therefore for any permutation $\alpha\in
\mathfrak S_N$, one has that

\begin{align}\label{e:alpha}
&(1-N)S(\rho)+\sum\limits_{m=1}^{N} H(M_{m})\notag\\
\geqslant&I(\alpha(1), \alpha(2), \cdots, \alpha(N)),
\end{align}
Taking the average of all permutations, we arrive at the following relation
\begin{align}\label{e:a}
&(1-N)S(\rho)+\sum\limits_{m=1}^{N} H(M_{m})\notag\\
\geqslant&\frac{\sum_{\alpha\in
\mathfrak S_N} I(\alpha(1), \cdots, \alpha(N))}{N!}.
\end{align}

Further analysis of the action of the symmetric group on the bound $I(\alpha(1), \cdots, \alpha(N))$ shows that only the first and the last
indices matter in the formula, as the bound is invariant under the action of any permutation from $\mathfrak S_{2, \cdots, N-1}$.
Among the remaining $N(N-1)$ permutations, it is enough to consider the cyclic group of $N$ permutations. Therefore the above average
can be simplified to the following form:
\begin{align}\label{e:a2}
&(1-N)S(\rho)+\sum\limits_{m=1}^{N} H(M_{m})\notag\\
\geqslant&\frac{\sum_{\mathrm{cyclic} \ \alpha} I(\alpha(1), \cdots, \alpha(N))}{N},
\end{align}
where the sum runs through all $N$ cyclic permutations $(12\cdots N), (23\cdots 1), \cdots,
(N1\cdots N-1)$.

Let's consider the case of three measurements $M_{m}$ in detail.
By using Eq. (\ref{e:basic}), we get that
\begin{align}
&-2S(\rho)+\sum\limits_{m=1}^{3} H(M_{m})\geqslant\notag\\
&-\alpha(\sum\limits_{i_{3}}p^{3}_{i_{3}}\log \sum\limits_{i_{1}, i_{2}} p^{1}_{i_{1}}c_{i_1i_2i_3}^{123})
:=\alpha(I(1, 2, 3)),
\end{align}
for any $\alpha\in\mathfrak {S}_{3}$, thus
\begin{align}
&-2S(\rho)+\sum\limits_{m=1}^{3} H(M_{m})\notag\\
\geqslant&\frac{1}{3}(I(1, 2, 3)+I(2, 3, 1)+I(3, 1, 2))\notag\\
=&\frac{\sum\limits_{i_{1}, i_{2}, i_{3}}p^{1}_{i_{1}}p^{2}_{i_{2}}p^{3}_{i_{3}}\log\sum p^{1}_{k_{1}}p^{2}_{k_{2}}p^{3}_{k_{3}}c_{k_1j_2 i_3}^{123}c_{k_2j_3i_1}^{231}c_{k_3j_1i_2}^{312}}{-3}£¬
\end{align}
where the sum inside logarithm runs over $j_1, j_2, j_3$, $k_1, k_2, k_3$.
For multi-index $(i_{1}, i_{2}, i_{3})$ we define the $d^{3}$-dimensional vector $\mathfrak{A}_{i_{1}, i_{2}, i_{3}}$ given by the elements
\begin{align}
\sum_{j_1, j_2, j_3}c_{k_1j_2 i_3}^{123}c_{k_2j_3i_1}^{231}c_{k_3j_1i_2}^{312},
\end{align}
and sorted in decreasing order with respect to
 multi-indices $(k_{1}, k_{2}, k_{3})$ (lexicographic order). Combined with the majorization bound $\omega\in \mathbb R^{d^3}$ formulated in section II,
 we immediately get that
\begin{align}\label{e:a123}
&-\log\sum p^{1}_{k_{1}}p^{2}_{k_{2}}p^{3}_{k_{3}}c_{k_1j_2 i_3}^{123}c_{k_2j_3i_1}^{231}c_{k_3j_1i_2}^{312}\notag\\
\geqslant&-\log(\omega\cdot\mathfrak{A}_{i_{1}, i_{2}, i_{3}}).
\end{align}
Then we introduce another $d^3$-dimensional vector $\mathfrak{B}$ defined by
$\mathfrak{B}_{i_{1}, i_{2}, i_{3}}=\log(\omega\cdot\mathfrak{A}_{i_{1}, i_{2}, i_{3}})$
and sorted in decreasing order with respect to multi-indices
$(i_{1}, i_{2}, i_{3})$ in the lexicographic order. Therefore we obtain the following {\it admixture} bound for 3 measurements
\begin{align}
-2S(\rho)+\sum\limits_{m=1}^{3} H(M_{m})\geqslant-\frac{1}{3}\omega\mathfrak{B}.
\end{align}
The new bound provides an improved lower bound for the uncertainty relation. In Fig. 3 we give an example
 to show that the admixture bound completely outperforms
 the other bounds that we have known  so far for multi-measurements. Moreover, this admixture bound
 can be easily extended to multi-measurements.

\begin{figure}
\centering
\includegraphics[width=0.4\textwidth]{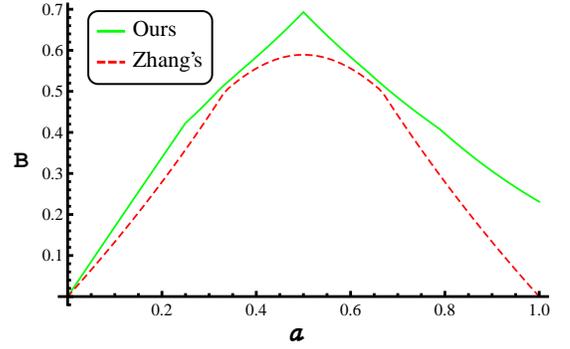}
\caption{Comparison of the admixture bound with
Zhang {\it et al.}'s bound for $\phi=\pi/2$ with $a\in [0, 1]$. Our bound in green is shown as the top curve
and always tighter.  Here $\ln$ is used on the bound axis $(B)$.}
\end{figure}

Let $M_{i}=\{|u^{i}_{i_{j}}\rangle\}$ be $N$ measurements, where $i=1, 2, \cdots, N$ and $j=1, 2, \cdots, d$.
 For each multi-index $(i_{1}, i_{2}, \cdots, i_{N})$ we introduce a $d^{N}$-dimensional vector
 $\mathfrak{A}_{i_{1}, i_{2}, \cdots, i_{N}}$ with the
  entries
  $$\sum_{\hat{\mathbf i}\mathbf j\cdots\hat{\mathbf k}}c^{12\cdots N}_{k_1j_2\cdots i_N}c^{23\cdots 1}_{k_2j_3\cdots i_1}
  \cdots c^{N1\cdots N-1}_{k_Nj_1\cdots i_{N-1}}
  $$
  where the sum runs over all indices except $\mathbf i=(i_1\cdots i_{N})$ and $\mathbf k=(k_1\cdots k_{N})$,
  and then sorted in decreasing order with respect to lexicographic order of multi-indices $(k_{1}, \cdots, k_{N})$. Set $log(\omega\cdot\mathfrak{A}_{i_{1}, i_{2}, \cdots, i_{N}}):=\mathfrak{B}_{i_{1}, i_{2}, \cdots, i_{N}}$ as the next
  $d^N$-dimensional vector with
  $\omega$ being the majorization bound for $N$ measurements formulated in the section \uppercase\expandafter{\romannumeral2}.
  Here $\mathfrak{B}_{i_{1}, i_{2}, \cdots, i_{N}}$ is assumed to be arranged in decreasing order
  with respect to the multi-indices $(i_{1}, i_{2}, \cdots, i_{N})$ lexicographically. The following result is
  then proved similarly as before.

\noindent\textbf{Theorem 3.} \label{t:main}
{\it The following entropic uncertainty relation holds,
\begin{align}
\sum\limits_{m=1}^{N}H(M_{m})+(1-N)S(\rho)\geqslant-\frac{1}{N}\omega\mathfrak{B}.  
\end{align}
}

The admixture bound is tighter than the previously known bounds. In fact, Fig. 2 depicts a comparison of our bound with that of
J. Zhang {\it et al.} \cite{Zhang}, while the latter is known to be tighter than the bound appeared in \cite{Fan}.

\section{\uppercase\expandafter{\romannumeral4}. Discussion}

In this paper, we have derived several tighter bounds for entropic uncertainty relations of multi-measurements
and in particular an admixture bound is obtained and proved to be tighter than all previously known bounds.
Inspired by the recent work
\cite{Coles,Fan,Friedland,Puchala,Rudnicki} we have taken the advantage of unitary matrix $U(S_{1},S_{2}, \cdots, S_{N})$
and come up with the universal bound for the multi-tensor products of distribution vectors.
To derive a deeper and better bound for $N$ measurements, we have studied the action of the symmetric group $\mathfrak S_N$
in combination with the universal vector bound of the distribution vectors and quantum channels. The derived admixture bound
turns out to be non-trivial bound for the uncertainties of $N$ measurements. Detailed comparisons with
previously known bounds are given in figures, and our admixture bound seems to outperform the other bounds most of the time.

Entropy characterizes and quantifies the physical essence of information resources in a mathematical manner.
The computational and operational properties of entropy make entropic uncertainty relations useful for quantum key distributions and other quantum
cryptography tasks, which can be performed relatively easy in a physical laboratory. Our new bounds are expected to be useful in handling
large data for these and further quantum information processings.

\smallskip
\noindent{\bf Acknowledgments}\, \, The work is supported by
NSFC (grant Nos. 11271138, 11531004), CSC and Simons Foundation grant 198129.

\end{document}